\pdfoutput=1
\documentclass[11pt]{article}
\usepackage{amssymb}
\usepackage{graphicx}
\parskip \baselineskip

\newtheorem{lm}{Lemma}

\newtheorem{prop}{Proposition}

\newtheorem{corol}{Corollary}

\newtheorem{ex}{Example}

\newtheorem{re}{Remark}

\newcommand{\ds}{\displaystyle}

\newcommand{\N}{\mathbb{N}}
\newcommand{\Nv}{{\bf Ner}}

\newcommand{\ol}{\overline}
\newcommand{\ul}{\underline}
\newcommand{\ra}{\rightarrow}

\date{}

\begin{document}

\title{Inclusion-exclusion enhanced by nerve stimulation}

\author{Marcel Wild}

\maketitle

\begin{quote}
{\bf Abstract} When evaluating the lengthy inclusion-exclusion expansion $N(\ul{\phi}) - N(\ul{1}) - N(\ul{2}) - \cdots + N(\ul{1}, \ul{2}) + N(\ul{1}, \ul{3}) + \cdots$ many of the terms $N(\cdots)$ may turn out to be zero, and hence should be discarded {\it beforehand}. Often this can be done. The main idea is that the index sets of nonzero terms constitute a set ideal (called the {\it nerve}) which can be encoded in a compact way. As a further enhancement, {\it equal} nonzero terms can sometimes be efficiently collected.
\end{quote}

\section{Introduction}

 Let $C(1), C(2) , \ldots, C(h)$ be constraints applying to a universe $\cal U$ of fixed objects. The reader may prefer to think of the $C(i)$'s as just subsets of $\cal U$. Let $N$ be the number of elements of $\cal U$ satisfying  {\it all} constraints, and e.g. write $N(\ul{2}, \ul{4})$ (or $N(\ul{4}, \ul{2})$) for the number of objects {\it violating} $C(2)$ and $C(4)$. One version of inclusion-exclusion (IE) states that

(1) \qquad  $N = |{\cal U}| - \Sigma N(\ul{i}) + \Sigma N(\ul{i}_1, \ul{i}_2) - \Sigma N(\ul{i}_1, \ul{i}_2, \ul{i}_3) - \cdots +(-1)^h N(\ul{1}, \ul{2}, \cdots, \ul{h})$

where say the third sum is taken over all ${h \choose 3}$ triplets $(i_1, i_2, i_3)$ with $1 \leq i_1< i_2 < i_3 \leq h$. 
In the sequel the universe ${\cal U}$ will be rather irrelevant. All that matters are the {\it index subsets} $\{i_1,\ldots,i_k\}$ of $[h]:=\{1,2,\ldots,h\}$ and their coupled numerical values $N(\ul{i_1},\ldots,\ul{i_k})$. For all index sets $X\subseteq [h]$, say $X=\{i,j,\ldots,k\}$, we often write $N(\ul{X})$ instead of $N(\ul{i}, \ul{j},\ldots, \ul{k})$. In particular $N(\ul{\emptyset})=|\cal U|$.

Our main aim is to compress the classic expansion (1) as 

(2) \qquad $N=\ds\sum_{k=0}^s N_k$

in various ways, all of which having in common that $N_0:=N(\ul{\emptyset})=|\cal U|$ and that each $N_k$ is a {\it sum of terms} $\pm N(...)$ in (1) such that terms occuring in $N_j$ and $N_k$ are distinct when $j\neq k$. One way
 (called {\it Upgrade B} like {\it B}asic) simply discards all {\it zero terms} $N(...)$ occuring in (1). Notice though that 'simply' is   an understatement. A refinement (called  {\it Upgrade A} like {\it A}dvanced) of Upgrade B is obtained by allowing $N_k$ in (2) to be a sum of several nonzero terms in (1), but all of
{\it equal magnitude.} Thus say $N_k=3+3-3+3+3-3$. In the remainder of the introduction we provide more details about Upgrade B and A.

 Let ${\cal P}[h]$ be the powerset of $[h]$. Call $X\subseteq {\cal P}[h]$ a {\it zeroset} if $N(\ul{X})=0$. Because from $X\subseteq Y\subseteq [h]$ and $N(\ul{X})=0$ follows $N(\ul{Y})=0$, the family  of zerosets  constitutes a set filter, which we call the {\it zeroset-filter} $\cal F$. Hence the
set system $\Nv:={\cal P}[h]\setminus{\cal F}$  is a set ideal or, synonimous, simplicial complex. It probably started in [NW] that in Combinatorial Topology the simplicial complex  $\Nv$ is sometimes called the {\it nerve} of the underlying IE problem. We shall obtain the nerve as a disjoint union of subcubes $r_1, r_2,\ldots$  of ${\cal P}[h]$. Each subcube $r_i$ can be viewed as  a length $h$ vector $r$ with entries $0,1,2$ where $2$ is a don't-care symbol, and so we refer to $r_i$ as a {\it 012-row}. Our 012-rows generalize to certain {\it 012n-rows}. 

Upgrade B is achieved by processing each 012n-row $r_i$ as follows. For each $X\in r_i$ calculate $N(\ul{X})$ and add or subtract $N(\ul{X})$ according to the parity of $|X|$. Section 2 provides the details, and illustrates the procedure on the task to count all permutations of $n$ letters that avoid certain {\it forbidden words}.

There are two types of Upgrade A: Horizontal or Vertical. 

Horizontal Upgrade A (Section 3) is e.g. applicable if the IE problem is such that for each $X\in\Nv$ the contribution $N(\ul{X})$ is an invariant of the cardinality $|X|$. Again we illustrate by means of permutations, this time subject to {\it forbidden positions}. Some  generalizations of this classic theme are offered. In lesser detail (more of them in [W3]) we discuss another application for Horizontal Upgrade A, namely the enumeration of upper-bounded integer compositions.

As to Vertical Upgrade A (Section 4), in principle this always is applicable, but it is the more efficient the fewer {\it distinct} summands $N(\ul{X})$ arise in (1), i.e. the smaller the {\it spectrum} $\{N(\ul{X}):\ X\subseteq [h]\}$ of the IE problem is. We  apply Vertical Upgrade A to count the number of models of a Boolean function in conjunctive normal form. In this scenario the spectrum consists of powers of 2 and thus is rather small.

The article in front of you (AFY) improves upon [W3] by cutting  slack, increasing readibility, and foremost by introducing Vertical Upgrade A. Other than [W3] with its Upgrade B for Boolean CNFs, the AFY features no numerical experiments. But they are planned for an upcoming version and they concern Vertical Upgrade A (again for Boolean CNFs).

\section{Upgrade B: Permutations with forbidden subwords}

To fix ideas, say our objects are all $N (\ul{\emptyset}) = 9!$ permutations $\pi$ of $[9]$, such as $\pi=376158429$. Consider these $h =6$ constraints $C(1)$ to $C(6)$:

(3) \qquad $\neg 123, \quad \neg 923, \quad \neg 9541, \quad \neg 3716, \quad \neg 379, \quad \neg 649$

Thus e.g. $C(2)$ holds in a permutation $\pi$ if $923$ is {\it not} a subword of $\pi$. Consequently $\neg C(2)$ holds in a $\pi$ if $923$ {\it is}  a subword of $\pi$.
For instance $N(\ul{1}, \ul{3})> 0$ since say $\pi = 895412367$ contains $123$ and $9541$, and so $\pi$ is counted by $N(\ul{1},\ul{3})$. One checks that $N(\ul{1}, \ul{3}), N(\ul{1}, 5), N(\ul{3}, 5)> 0$ but $N(\ul{1}, \ul{3}, \ul{5})=0$ since the simultaneous occurence of $379$ and $9541$ (thus $379541$) and $123$ is impossible. Albeit a little tedious, one verifies ad hoc that the minimal zerosets $X\subseteq [h]$ are:

(4) \qquad $\{1, 2\}, \{1, 4\}, \{2, 3\}, \{2, 5\}, \{3, 4\}, \{3, 6\}, \{4, 5\}, \{5, 6\}, \{1, 3, 5\}, \{2, 4, 6\}.$

Thus the ten sets in (4) are the {\it generators} (=minimal members) of the zeroset-filter ${\cal F} \subseteq {\cal P}[h]$. Recall that $\Nv = {\cal P}[h] \backslash {\cal F}$.  As for any simplicial complex, we call the sets $U \in \Nv$  the {\it faces} of $\Nv$.
Feeding the generators $\Gamma_i$ of ${\cal F}$ to the $n$-algorithm of [W2] one obtains $\Nv = r_1 \uplus r_2 \uplus r_2 \uplus r_4$ as a disjoint union of set systems $r_1$ to $r_4$ as defined in Table 1:

\begin{tabular}{l|c|c|c|c|c|c|}
&  1 & 2& 3 & 4 & 5 & 6\\ \hline
$r_1 =$ & 1 & 0 & 0 & 0 & 0 & 2  \\ \hline
$r_2 =$ & 0 & $n$ & 0 & $n$ & 0 & $n$  \\ \hline
$r_3=$ & 2 & 0 & 1 & 0 & 0 & 0  \\ \hline
$r_4=$ & $n$ & 0 & $n$ & 0 & 1 & 0  \\ \hline
\end{tabular}

{\sl Table 1: Compressed representation of the nerve}

Each such 012n-{\it row} $r_i$ comprises a family of bitstrings $u$ whose supports $U \subseteq [6]$ are faces of $\Nv$.
Besides the {\it don't-care} symbol $2$ which can freely be chosen  0 or 1, we use the wildcard $n n \cdots n$ which means ``at least one $0$''. In other words, only $1 1 \cdots 1$ is forbidden. Thus say $r_2$ contains $2^3 -1 = 7$ bitstrings, one of them being $(0,0,{\bf 1}, 0, {\bf 1},0,{\bf 0})$ and matching the face $\{3,5\}$. Also the empty face $\emptyset$ is in $r_2$ (set $nnn=000$). Because the produced 012n-rows are mutually disjoint it follows that

(5) \quad $$|\Nv| =  |r_1|+\cdots+|r_4|=2 + 7 + 2 +3 = 14.$$

Scanning the 14 faces of $\Nv$ row-wise (indicated by the bracketings) yields

(6) \quad $N = \Big( N(\ul{1}, \ul{6}) - N(\ul{1}) \Big) \ + \  \Big( N(\ul{2}, \ul{4}) + N(\ul{2}, \ul{6}) 
+ N(\ul{4}, \ul{6})-N(\ul{2}) - N(\ul{4})-N(\ul{6}) + N(\ul{\emptyset})\Big)$

\hspace*{1.6cm} $+\Big( N(\ul{1}, \ul{3})- N(\ul{3}) \Big) \ + \ \Big(  N(\ul{1},\ul{5}) +N(\ul{3},\ul{5}) - N(\ul{5}) \Big)$.

This is a type (2) expansion with  $s+1=14$ summands $N_0,\ N_1,...,N_{13}$, thus less than the 64 summands in the type (1) expansion. Except for $N_0=N(\ul{\emptyset})=6!$ it does not matter how we assign the values $N_k$ in (2) to the summands in (6).

For instance the $N(\ul{1}, \ul{6})$ many permutations $\pi$ of [9] satisfying $\neg C(1) \wedge \neg C(6)$ in (3) match the permutations of the blocks $123, 649, 5, 7, 8$, and so $N(\ul{1}, \ul{6}) = 5!$. Likewise $N(\ul{4}, \ul{6}) = 4!$ is the number of permutations of $371649, 2, 5, 8$. 

 

{\bf 2.1} If all generators $\Gamma_i$ of the zeroset-filter ${\cal F}$ are 2-element, i.e. matching the edges of a graph $G$, then $\Nv$ consists of all anticliques (=independent sets) of $G$. Instead of feeding all edges $\Gamma_i$ of $G$ to the $n$-algorithm, it would be more economic if the fewer vertices of $G$ could be processed, somehow. In a nutshell, this is how to do it. Say $3 \in V(G)$ with set of neighbours $NB(3) = \{1, 4, 7\}$. If $X \subseteq V(G)$ is an anticlique that happens to contain 3, then $NB(3) \cap X = \emptyset$. In other words, each anticlique $X$ satisfies the ``anti-implication'' $3 \ra \ol{1} \wedge \ol{4} \wedge \ol{7}$, and similarly for the  other $h-1$ vertices $\neq 3$. Conversely, any set $X \subseteq V(G)$ satisfying these $h$ {\it anti-implications} necessarily is an anticlique. A symbolic notation for the family of all sets $Y \subseteq V(G)$ satisfying 
$3 \ra \ol{1} \wedge \ol{4} \wedge \ol{7}$ is $( c, 2, a, c, 2,2, c)$, assuming that $h=7$. Formally

(7) \quad $(c, 2, a, c, 2,2, c): = (2, 2, {\bf 0}, 2,2,2,2) \uplus (0, 2, {\bf 1}, 0, 2, 2, 0).$

This leads to the $ac$-algorithm of [W1] (see also Subsection 5.2.1) which represents the anticliques of any graph as a disjoint union of 012ac-rows.

{\bf 2.2} In summary, the {\it Upgrade B} of inclusion-exclusion is as follows.


\begin{enumerate}
	\item [(B)] Provided the generators $\Gamma_i$ of the zeroset-filter ${\cal F}\subseteq P[h]$ can be found with moderate effort, 
	one can represent the nerve as $\Nv= r_1 \uplus r_2 \uplus + \cdots \uplus r_R$ with multi-valued rows $r_i$ of length $h$.  Calculating $N(\ul{i}_1, \ldots, \ul{i}_k)$ for each face $\{i_1, \cdots, i_k\} \in r_j\ (1\le j\le R)$ one gets $N$ as  
	$N = \Sigma \{(-1)^k N(\ul{i}_1, \cdots, \ul{i}_k ): \ \{i_1, \cdots, i_k \} \in \Nv \}$, which matches pattern (2).
\end{enumerate}

Here ``multi-valued row '' means 012n-row or 012ac-row. Due to the overhead of Upgrade B classic IE may excel for small examples (say $|{\cal U}|<20$). Otherwise Upgrade B wins out, although this may mean 1 year versus a million years computation time. Whether Upgrade B is itself feasible depends to large extent on the time
 to compute the generators $\Gamma_i$.
Once the $\Gamma_i$'s are available, $|{\Nv}|$ can be predicted  by applying off-the-shelf algorithms (like Mathematica's {\tt SatisfiabilityCount}) to a Boolean
 formula readily derived  from the $\Gamma_i$'s. 
Upon knowing $|\Nv|$ one can decide whether Upgrade B (or A in Section 3) pays off or whether one should drop the IE endeavour altogether.
 As to the formal cost, the $n$-algorithm displays $\Nv$ as a disjoint union of $R$ many 012n-rows $r_1, r_2, \cdots, r_R$ in polynomial total time $O(Rm^2h^2)$ according to [W2] (respectively [W1] for 012ac-rows).




\section{Horizontal Upgrade A: Permutations with forbidden positions, respectively upper-bounded integer compositions}

While partitioning $\Nv \subseteq {\cal P}[h]$  into 012n-rows beats classic IE, $\Nv$ may still be too large to be scanned one by one. But sometimes one can cope as follows. Suppose for each face  $\{i_1, \cdots, i_k\} \in \Nv$ the number $N(\ul{i}_1, \cdots, \ul{i}_k)$ is an invariant of $k$, thus 

(8)\quad $N(\ul{i}_1, \cdots, \ul{i}_k) = g(k)$ for some function $g$ from $[h]\cup\{0\}$ to $\N$. 

(So $N(\ul{\emptyset})=g(0)$.) Then it pays to calculate the {\it face numbers}

$f(k): = |\{U \in \Nv: \ |U|=k\}| \quad (0 \leq k \leq h)$

in order  to calculate $N$ with type (2) compression as

(9) \qquad $N = \ds\sum_{k=0}^h (-1)^k f(k) g(k)$.

Thus $N_k$ in (2) is $(-1)^k f(k)g(k)$. In particular $N_0=f(0)\cdot g(0)=1\cdot N(\ul{\emptyset})$, as it must. 
In the remainder of Section 3 we show how  {\it Horizontal Upgrade A}  applies to counting permutations constrained in novel ways. Specifically, after reviewing the classic problem of ``forbidden positions'' (3.1), this gets generalized (3.2) to the scenario where positions occupied by certain letters force other positions to be avoided by certain letters. In 3.3 we further generalize permutations to  arbitrary, injective, or surjective maps respectively.
Subsection 3.4 shows how Horizontal Upgrade A counts upper-bounded integer compositions.
Our examples are small enough for the face numbers $f(k)$ required in (9) to be found by inspection. How this is done for Horizontal Upgrade A in general is shown in Section 4.

{\bf 3.1} We wish to count the permutations $\pi$ of [10] that satisfy this conjunction $C'(1)\wedge\ldots\wedge C'(6)$:

(10) \qquad $\pi (3) \neq 1  \quad \wedge \quad \pi (1) \neq 5  \quad \wedge \quad \pi (2) \neq 1 
      \quad \wedge \quad \pi (6) \neq  4 \quad \wedge \quad \pi (1) \neq 3 \quad \wedge \quad \pi (2) \neq 7$

If again we write permutations as words this amounts to the familiar problem of counting permutations with forbidden positions, thus 3 and 5 must not be at the beginning, 1 and 7 not at position 2, and so forth. This classic problem is often viewed as placing non-taking rooks on mutilated chessboards. We rather view it (and generalize it in 3.2) as  the problem to find all anticliques of a suitable graph.


{\bf 3.2} The upcoming generalization seems to be new.
Namely, the constraints $C'(i)$ in (10) get weakened to  disjunctions $C(i)$ as follows.


$\begin{array}{llllll}
C(1): & \pi (3) \neq 1 & \vee & \pi (4) \neq 2 & \vee & \pi (5) \neq 3 \\
\\
C(2): & \pi (1) \neq 5 & \vee & \pi (4) \neq 6 & \vee & \pi (5) \neq 4\\
\\
C(3) : & \pi (2) \neq 1 & \vee & \pi (6) \neq 7 & \vee & \pi (7) \neq 10 \\
\\
C(4) : & \pi (6) \neq 4 & \vee & \pi (8) \neq 6 & \vee & \pi (10) \neq 7 \\
\\
C(5) : & \pi (1) \neq 3 & \vee & \pi (8) \neq 5 & \vee & \pi (9) \neq 8\\
\\
C(6) : & \pi (2) \neq 7 & \vee & \pi (7) \neq 2 & \vee & \pi (10) \neq 10 \end{array}$  \hfill (11)

The six constraints $C'(i)$ from (10) match the first column in the display (11). Of course say
 
$$\neg C(1) \quad \mbox{means} \quad \pi (3) = 1 \quad \wedge \quad \pi (4) = 2 \quad \wedge \quad \pi (5) = 3.$$

So for instance $\neg C(4) \ \wedge \ \neg C(6)$ entails the clash $(\pi (10) = 7 \ \wedge \ \pi (10) = 10)$, which contradicts $\pi$ being a function. Hence $N(\ul{4}, \ul{6}) =0$, hence $\{4,6\} \in {\cal F}$. Similarly $\{1, 3\} \in {\cal F}$ since $\neg C(1) \ \wedge \ \neg C(3)$ entails $(\pi (3) = 1 \ \wedge \ \pi (2) = 1)$ which contradicts  the injectivity of $\pi$. Obviously $\{4,6\}$ and $\{1,3\}$ are minimal members, i.e. generators of ${\cal F}$.

{\bf 3.2.1} We claim that  {\it all} generators of ${\cal F}$ are 2-element. 
Indeed, if $N(\ul{i}_1, \ldots, \ul{i}_k) = 0$, i.e. if $\neg C(i_1) \wedge \cdots \wedge \neg C(i_k)$ has $0$ models, then this must\footnote{Proof by contraposition: Suppose that $\neg C(i_1) \wedge \cdots \wedge \neg C(i_k)$ expands as a conjunction of identities $\pi(i)=j,\ \pi(k)=\ell,\ \ldots$ of the following kind.We never demand an index to be mapped to distinct indices, and never demand different indices to map to the same index. Then evidently there exists at least one permutation $\pi$ satisfying all identities.}  be caused by one or more clashes as above.
 It follows that the generators of $\cal F$ match the edges of a graph $G$ (see 2.1). Ad hoc checking all 2-element subsets of [6] yields the graph $G$ in Figure 1 (the dashed edges count as well). It follows that $\Nv={\cal P}[6]\setminus {\cal F}$ is the simplicial complex of all anticliques of $G$. Its face-numbers $f(k)$ are the numbers of $k$-element anticliques of $G$. One finds by inspection

 $$f(1) = 6, \ f(2) = 7, \ f(3) = 1\ ({\it i.e.}\ \{1,5,6\}), \ f(4) = f(5) = f(6) =0.$$

\begin{center}
\includegraphics[scale=0.5]{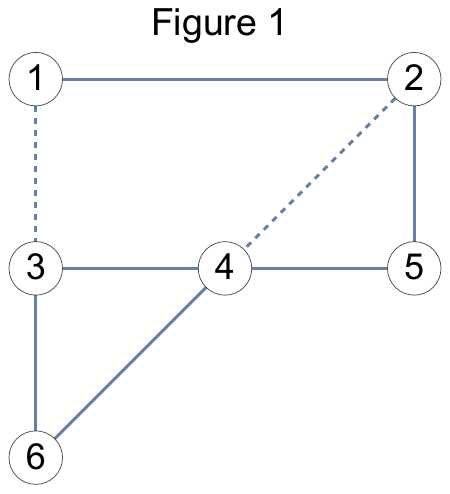}
\end{center}

{\bf 3.2.2} Let us show that $N(\ul{i}_1, \cdots, \ul{i}_k)$ (if nonzero) is an invariant $g(k)$.
For starters, we claim there are $N(\ul{5},\ul{6}) = 4!$ permutations $\pi$ of [10] violating $C(5)$ and $C(6)$, i.e. satisfying 

$$\Big(\pi (1) = 3 \ \wedge \ \pi (8) = 5 \ \wedge \ \pi (9) = 8\Big) \ \wedge \ 
\Big(\pi (2) = 7 \ \wedge \ \pi (7) = 2 \ \wedge \ \pi (10)\Big) = 10.$$

Here $4!$ arises as $(10 - 3 \cdot 2)!$ due to the mere fact that $\{1, 8, 9\}$ and $\{2, 7, 10\}$ are disjoint. A glance at (11) confirms that incidently\footnote{ This kind of {\it Purity Condition} may or may not hold. If it does not hold then only Upgrade B applies.} each constraint
$\pi (x) =y$ occurs in {\it at most one} of $\neg C(1), \cdots, \neg C(6)$; for instance $\pi (5) = 4$ only occurs in $\neg C(2)$. It follows that generally

$$N(\ul{i}_1, \cdots, \ul{i}_k) = (10 - 3k)! = : g(k).$$

Therefore Horizontal Upgrade A (see (9)) yields

 $$N =  \ds\sum_{k=0}^6 (-1)^k f(k) g(k) = 10! - 6 \cdot 7! + 7 \cdot 4! - 1 \cdot 1!+0-0+0 =3598727.$$

The disjunction, say, $C(1)$ can also be viewed as {\it implication} $(\pi (3) = 1 \wedge \pi (4) =2) \ra (\pi (5) \neq 3)$, or equivalently as the implication $(\pi (3) = 1 \wedge \pi (5) = 3) \ra (\pi (4) \neq 2)$ in (11). Other than in (11), as long as the Purity Condition holds, each $C(k)$ may consist of any number of inequalities $\pi(i)\neq j$.

{\bf 3.3} Instead of permutations  let us count  arbitrary maps $\pi : [10] \ra [10]$ that satisfy the six constraints Observe that now $N (\ul{1}, \ul{3}) > 0$ since $(\pi (3) = 1$ and $\pi (2) = 1$) is  allowed. In terms of the graph in  Figure 1, there is no longer an edge between $1$ and $3$. Similarly the edge between $2$ and $4$ disappears. Hence the nerve is the simplicial complex $\Nv \subseteq {\cal P}[6]$ of all anticliques of the adjusted graph omitting the dashed edges. By inspection one finds that its face numbers are $f(1) =6, \ f(2) = 9, \ f(3) = 2, \ f(4) = f(5) = f(6) = 0$. As opposed to $g(k) = (10-3k)!$ in 3.2.2 here
 $g(k) = 10^{10-3k}$, and so

 $$N = \ds\sum_{k=0}^6 (-1)^k f(k) g(k) = 10^{10} - 6 \cdot 10^7 + 9 \cdot 10^4 - 2\cdot 10^1+0-0+0= 940089980.$$

 In a similar fashion injective or (using Stirling numbers) surjective maps $\pi:[10]\ra [10]$ constrained by (11) can be dealt with.

{\bf 3.4} Horizontal Upgrade A  can be generalized as follows. Instead of $N(\ul{i}_1, \cdots, \ul{i}_k )$ being a function of $k$
 as in (8), it suffices that $N(\ul{i}_1, \cdots, \ul{i}_k )$ is a function of any function {\it val} of $\{i_1, \ldots, i_k \}$. Thus

(8') \quad $N(\ul{i}_1, \cdots, \ul{i}_k )=g(\,{\it val}\{i_1, \ldots, i_k \}\,)$  for some function $g$.

Putting ${\it val}\{i_1, \ldots, i_k \}:=k$ formula (8') boils down to (8). 

As a proper new example consider the problem to calculate the number $N$
 of (non-negative) {\it integer compositions} $(u_1,\ldots,u_6)$ of 9, subject to the restrictions

(12) \quad  $u_i < a_i$ \ where $a_1 = 7, \, a_2 =4, \, a_3 = a_4 = 3, \ a_5 = a_6 =2$.


For any integer composition $(u_1,\ldots,u_6)$ of 9 let $C(i)$ be the property  that $u_i<a_i$. Accordingly, if $\{i_1, \ldots, i_k \}\subseteq [6]$ then $N(\ul{i}_1, \cdots, \ul{i}_k)$ is the number of integer compositions having 
$u_{i_1} \geq a_i, \cdots, u_{i_k} \geq a_{i_k}$. While $N(\ul{i}_1, \cdots, \ul{i}_k)$ is {\it not} determined by $k$, it is determined by the {\it value} of $\{i_1, \cdots, i_k\}$ if this is defined as

$$v=val\{i_1, \cdots, i_k\} : = \ a_{i_1} + \cdots + a_{i_k}.$$

Namely, one can show\footnote{For  details see [W3]. There our approach to counting restricted integer compositions is furthermore compared with other methods. A more thorough investigation, taking into account the magnitudes of $u_i,\ a_i\ (1\le i\le h)$ and $h$ itself, seems worthwile.} that

(13) \quad $N(\ul{i}_1, \cdots, \ul{i}_k) = g(v):=\ds {14-v \choose 5} $.

For instance  $\{i_1, \cdots, i_k\}=\emptyset$ yields $v=0$, and so $N(\ul{\emptyset})=g(0)={14\choose 5} $ is the number of (unrestricted) integer compositions of $9$. Likewise $\{i_1, \cdots, i_k\}=\{3,4,5\}$ yields $v=a_3+a_4+a_5=8$, and so the number of integer compositions of $9$ having $u_3\ge 3,\ u_4\ge 3,\ u_5\ge 2$ equals $N(\ul{3},\ul{4},\ul{5})=g(8)=6$. One checks
 ad hoc that the six integer compositions are the ones in Table 2.

\begin{tabular}{l|c|c|c|c|c|c|}
$u_1$ & $u_2$ & $u_3$ & $u_4$ & $u_5$ & $u_6$\\ \hline
1 & 0 & 3 & 3 & 2 & 0 \\ \hline 
0 & 1 & 3 & 3 & 2 & 0 \\ \hline
0 & 0 & 3 & 3 & 2 & 1 \\ \hline
0 & 0 & 4 & 3 & 2 & 0 \\ \hline
0 & 0 & 3 & 4 & 2 & 0 \\ \hline
0 & 0 & 3 & 3 & 3 & 0 \\ \hline \end{tabular}

{\sl Table 2: The admissible number compositions of 9}

\section{Horizontal Upgrade A: Systematic calculation of the face numbers}

 Once the nerve of an IE problem has been obtained as a disjoint union of 012n-rows or 012ac-rows, i.e. $\Nv = r_1 \uplus r_2 \uplus + \cdots \uplus r_R$, then the face-numbers can be obtained as 

(14)\quad  $f(k) = \, \mbox{Card}(r_1, k) + \cdots + \, \mbox{Card}(r_R, k),$ 

where $\mbox{Card}(r, k) : = |\{U \in r: |U| = k \}|.$
Other than in Table 1, a general 012n-row $r$ can have {\it several} $n$-wildcards, which are then distinguished by subscripts. In order to calculate all numbers Card$(r, k)$ for say
$$r: = (0,1,2,2,2,\, n_1, n_1,\, n_2, n_2,\, n_3, n_3, n_3, n_3, n_3)$$
we associate\footnote{Article [W4,p.124] first introduced this particular kind of counting polynomial in a dual context ($e$-algorithm instead of $n$-algorithm). The formal cost of expanding products of polynomials is also investigated there.} with each component 1 of $r$ the polynomial $x$, with each component $2$ the polynomial $1+x$, and with each $n$-wildcard $(n,n, \cdots, n)$ of length $t$ the polynomial $(1+x)^t-x^t=1+tx + {t \choose 2} x^2 + \cdots {t \choose t-1} x^{t-1}$, and multiply out. For $r$ as above this results in

\begin{description}
\item{(15)}\quad $p(x) =  x \cdot (1+x)^3 \cdot (1+2x)^2 \cdot (1+5x + 10x^2 + 10x^3+5x^4) $
\item{} \hspace{1.8cm} $=x + 12x^2 + 64x^3 + 200x^4 + 406x^5+ 559x^6 + 525x^7+ 325x^8 + 120x^9 + 20x^{10}.$
\end{description}

It is not hard to see (and is a standard technique in enumerative combinatorics) that always the coefficients of the expanded polynomial yield the sought numbers Card$(r,k)$; say Card$(r,5) = 406$. 
The Mathematica command {\tt Expand$[p[x]]$} readily does the job. Alluding to the 'horizontal' 012n-rows the described method will be called
{\it Horizontal Upgrade A} (H-A) of inclusion-exclusion. To summarize:

\begin{enumerate}
	\item [(H-A)] Provided the generators of the zeroset-filter ${\cal F}\subseteq P[h]$ can be found with moderate effort, 
	one can represent the nerve as $\Nv= r_1 \uplus r_2 \uplus + \cdots \uplus r_R$ with 012n-valued rows $r_i$ of length $h$. 
	Using (14) and (15) the face-numbers $f(k)$ of $\Nv$ can be calculated fast. If the invariance property (8) holds, this can be exploited to get $N$ as
	 $N = \sum_{k=0}^h (-1)^k f(k) g(k)$, which matches pattern (2).
\end{enumerate}

\section{Vertical Upgrade A: Counting the models of a Boolean CNF}

In 5.1 we outline the general features of Vertical Upgrade A. In 5.2 this machinery is applied to counting the models of a Boolean CNF. Some technical details are deferred to 5.3.

{\bf 5.1} Suppose the nonzero values $N(\ul{X})$ of an IE problem based on $\cal U$  are $v_1<v_2<\cdots<v_t$. If for all $1\le j\le t$ we define

\begin{description}
\item{(16) } $N[v_k]':=|\{X\in\Nv:\ N(\ul{X})=v_k\ {\it and}\ |X|\ \hbox{\it is odd}\}|$    
\item{} \hspace{0.8cm} $N[v_k]'':=|\{X\in\Nv:\ N(\ul{X})=v_k\ {\it and}\ |X|\ \hbox{\it is even}\}|$
\end{description}

then we obtain this type (2) compression with $s=2t$:

(17)\quad $N= \ds \sum_{k=1}^t v_k N[v_k]'' -\ds \sum_{k=1}^t v_k N[v_k]'$

Here $N_0=|{\cal U}|$ appears as $N[v_t]''$. In order to calculate the numbers $N[v_k]'$ and $N[v_k]''$ observe that each set system 

$\Nv[\ge v_i]:=\{X\in \Nv:\ N(\ul{X})\ge v_i\}$

is a nonempty simplicial complex. Obviously $\Nv[\ge v_1]=\Nv$ and

(18)\quad $\Nv[\ge v_1]\supset \Nv[\ge v_2]\supset \cdots \supset \Nv[\ge v_t]$

is a  {\it filtration} of $\Nv$ with strict inclusions. We put

\begin{description}
\item{(19) } $\Nv[\ge v_k]':=\{X\in \Nv[\ge v_k]:\ |X|\ \hbox{\it is odd} \}$,     
\item{} \hspace{0.8cm} $\Nv[\ge v_k]'':=\{X\in \Nv[\ge v_k]:\ |X|\ \hbox{\it is even} \}$.    
\end{description}

It is evident that for all $(1\le k<t)$ one has

\begin{description}
\item{(20) } $|\Nv[\ge v_k]'| -  |\Nv[\ge v_{k+1}]'| = N[v_k]',$
\item{} \hspace{0.8cm}   $|\Nv[\ge v_k]''| -  |\Nv[\ge v_{k+1}]''| = N[v_k]''.$
\end{description}

For $k=t$ formula (20) does not apply. But in this case $N[v_t]''=|{\cal U}|$ and $N[v_t]'=0$. 
In principle formula (17), which we henceforth call\footnote{In computer implementations $\Nv$ is given as disjoint union of 012n-rows. The adjective 'vertical' derives from the fact that each simplicial comples $\Nv[\ge v_i]$ cuts several 'horizontal' 012n-rows.} {\it Vertical Upgrade A}, applies to every IE problem. But it is useful only when the cardinalities on the left in (20) can be obtained smoothly.

{\bf 5.2} Let us apply the theory in 5.1 to count the models of a Boolean function $\varphi(x_1,\ldots,x_n)$ given as  conjunctive normal form (CNF), i.e. as a conjunction of $h$ disjunctions (=clauses). In the example below we  have $n=h=6$ (whereas in most applications $h>n$).

(21)\quad $\varphi(x_1,\ldots,x_6):=(x_1\vee\ol{x_2}\vee x_3)\wedge(x_1\vee x_3\vee\ol{x_6})\wedge(\ol{x_1}\vee x_3)\wedge
(\ol{x_4}\vee x_6)\wedge (\ol{x_2}\vee \ol{x_3})\wedge x_4$

We say that a bitstring ${\bf x}\in{\cal U}=\{0,1\}^n$ has property $C(i)$ if {\bf x} satisfies the $i$-th clause in (21). Consequently {\bf x} satisfies  $\neg C(2)$ if $\ol{x_1}\wedge\ol{x_3}\wedge x_6=1$. Generally for all $i\in [h]$ the set of bitstrings $T_i\subseteq\{0,1\}^n$ satisfying the conjunction (=term) $\neg C(i)$ is given as 012-row in Table 3.

\begin{tabular}{l|c|c|c|c|c|c|}
&  1 & 2& 3 & 4 & 5 & 6\\ \hline
$T_1 =$ & 0 & 1 & 0 & 2 & 2 & 2  \\ \hline
$T_2 =$ & 0 & 2 & 0 & 2 & 2 & 1  \\ \hline
$T_3=$ & 1 & 2 & 0 & 2 & 2 & 2  \\ \hline
$T_4=$ & 2 & 2 & 2 & 1 & 2 & 0  \\ \hline
$T_5=$ & 2 & 1 & 1 & 2 & 2 & 2  \\ \hline
$T_6=$ & 2 & 2 & 2 & 0 & 2 & 2  \\ \hline
\end{tabular}

{\sl Table 3: Compressed representation of $\{0,1\}^n\setminus Mod(\varphi)$}

We conclude that

(22)\quad $N(\ul{i_1},\ldots,\ul{i_k})=|T_{i_1}\cap\cdots\cap T_{i_k}|  $

for all index subsets $\{i_1,\ldots,i_k\}\subseteq [h]$. The intersection $D$ of any number of 012-rows of length $n$ is easily determined componentwise according to the rules $a\wedge b=b\wedge a\ (a,b\in\{0,1,2\})$ and

$$0\wedge 2=0\wedge 0=0,\quad 1\wedge 2=1\wedge 1=1,\quad 2\wedge 2=2.$$

 Thus say $D=T_1\cap T_2\cap T_6=(0,1,0,0,2,1)$ where the last $1$ is obtained as $2\wedge 1\wedge 2=1$. If at some position $0$ clashes with $1$ then $D=\emptyset$. So $T_4\cap T_6=\emptyset$ because of a clash at the fourth position. Generally
$T_{i_1}\cap\cdots\cap T_{i_k}=\emptyset$ entails that $T_\alpha\cap T_\beta=\emptyset$ for some $\alpha,\ \beta\in\{i_1,\ldots,i_k\}$. Accordingly 
the zeroset-filter is generated by the edges $\{\alpha,\beta\}$ of a graph $G=(V,E)$. This makes $\Nv$ the set of anticliques of $G$ (see 2.1). In our example $G=([6],E)$ is rendered in Figure 2.

\begin{center}
\includegraphics[scale=0.5]{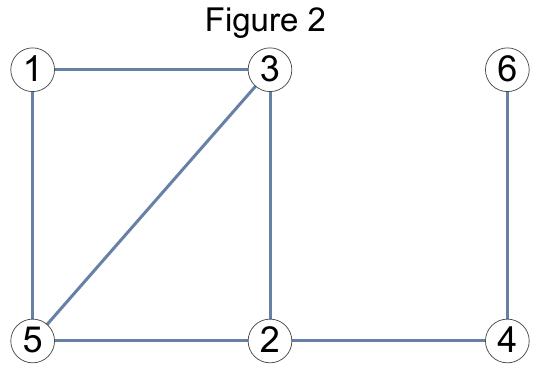}
\end{center}

{\bf 5.2.1} In order to represent $\Nv$ as a disjoint union of $012ac$-rows we delve a bit deeper into the ac-algorithm of Section 2.1.

\begin{tabular}{l|c|c|c|c|c|c|}
&  1 & 2& 3 & 4 & 5 & 6\\ \hline
$\rho_1 =$ & c & c & a & 2 & c & 2  \\ \hline
$\rho_2 =$ & 0 & 0 & 0 & 2 & 1 & 2  \\ \hline
$\rho_3=$ & c & c & a & 2 & 0 & 2  \\ \hline
$\rho_4=$ & 0 & 0 & 0 & a & 1 & c  \\ \hline
$\rho_5=$ & c & 0 & a & {\bf 1} & 0 & 0  \\ \hline
$\rho_6=$ & c & c & a & {\bf 0} & 0 & 2  \\ \hline
\end{tabular}

{\sl Table 4: The workings of the ac-algorithm}

Recall that all $h$ vertices of $G$ give rise to  anti-implications. The latter need to be imposed on 
shrinking set systems, the first one being  ${\cal U}=\{0,1\}^n=(2,2,2,2,2,2)$. Actually not all $h$ anti-implications are necessary. It suffices to take the ones that correspond to the vertices in any fixed edge-cover $S\subseteq V$. Here we take $S=\{3,5,4\}$ (the ordering is due to convenience of presentation), which thus yields the anti-implications

$$3 \ra \ol{1} \wedge \ol{2} \wedge \ol{5},\quad  5 \ra \ol{1} \wedge \ol{2} \wedge \ol{3},\quad 4 \ra \ol{2} \wedge \ol{6}. $$

Imposing the first anti-implication upon $(2,2,2,2,2,2)$ yields row $\rho_1$ in Table 4. In order to impose $5 \ra \ol{1} \wedge \ol{2} \wedge \ol{3}$ upon $\rho_1$ we write

$$\rho_1=\rho_2'\uplus\rho_3:=(2,2,0,2,{\bf 1},2)\uplus(c,c,a,2,{\bf 0},2).$$

It follows that $\rho_2\uplus\rho_3$ (see Table 4) contains exactly those $X\in\rho_1$ that satisfy $5 \ra \ol{1} \wedge \ol{2} \wedge \ol{3}$. Clearly the sets $X\in\rho_2$ satisfying $4 \ra \ol{2} \wedge \ol{6}$ are exactly the sets in $\rho_4$. A moment's thought shows that the sets $X\in\rho_3$ satisfying $4 \ra \ol{2} \wedge \ol{6}$ are exactly the sets in $\rho_5\uplus\rho_6$. To summarize,

(23)\quad $\Nv=\rho_4\uplus\rho_5\uplus\rho_6.$

Using (23) we could proceed with Upgrade B to calculate $N=|Mod(\varphi)|$. In fact in [W3] a Mathematica implementation of Upgrade B was applied to random Boolean functions of various shape.

{\bf 5.2.2.} Yet here we indicate how Upgrade B can be pushed to Vertical Upgrade A. Sticking to the example initiated in 5.2.1 the filtration of $\Nv$ in (18) will be obtained by applying the $n$-algorithm to the $012ac$-rows $\rho_4,\ \rho_5,\ \rho_6$ in (23). For simplicity let us first\footnote{In the planned Mathematica implementation of Vertical Upgrade A this translation will not be necessary since an adaption of the $n$-algorithm can digest raw $012ac$-rows. Such an adaption likely pays off since translating $012ac$-rows into $012n$-rows increases the number of rows. (Albeit merely from 3 to 4 in our toy example.)} translate $\rho_4,\ \rho_5,\ \rho_6$ to $012n$-rows $r_i$ as follows (see Table 5): $r_1=\rho_4,\ r_2=\rho_5,\ r_3\uplus r_4=\rho_6$.

\begin{tabular}{l|c|c|c|c|c|c|c}
&  1 & 2& 3 & 4 & 5 & 6\\ \hline
$r_1 =$ & 0 & 0 & 0 & $n$ & 1 & $n$ & $\ra 1+2$ \\ \hline
$r_2 =$ & $n$ & 0 & $n$ & 1 & 0 & 0 & $\ra 1+2$ \\ \hline
$r_3=$ & 2 & 2 & {\bf 0} & 0 & 0 & 2 & $\ra 4+4$ \\ \hline
$r_4=$ & 0 & 0 & {\bf 1} & 0 & 0 & 2 & $\ra 1+1$ \\ \hline
 &  &  &  &  &  &  & \\ \hline
$r_5=$ & $n$ & $n$ & 0 & 0 & 0 & $n$ & $\ra 3+4$ \\ \hline
$r_6 =$ & 0 & 0 & 1 & 0 & 0 & 2 & $\ra 1+1$ \\ \hline
$r_7 =$ & 0 & 0 & 0 & 0 & 1 & 2 & $\ra 1+1$ \\ \hline
$r_8=$ & 0 & 0 & 2 & 1 & 0 & 0 & $\ra 1+1$ \\ \hline
$r_{9}=$ & 0 & 0 & 0 & 1 & 1 & 0 & $\ra 0+1$ \\ \hline
&  &  &  &  &  &  & \\ \hline
$r_{10}=$ & 0 & $n$ & 0 & 0 & 0 & $n$ & $\ra 2+1$ \\ \hline
$r_{11} =$ & 1 & 0 & 0 & 0 & 0 & 0 & $\ra 1+0$ \\ \hline
$r_{12} =$ & 0 & 0 & 1 & 0 & 0 & 2 & $\ra 1+1$ \\ \hline
$r_{13}=$ & 0 & 0 & 0 & 0 & 1 & 2 & $\ra 1+1$ \\ \hline
$r_{14}=$ & 0 & 0 & 0 & 1 & 0 & 0 & $\ra 1+0$ \\ \hline
&  &  &  &  &  &  & \\ \hline
$r_{15}=$ & 0 & 0 & 0 & 0 & 0 & 2 & $\ra 1+1$ \\ \hline
$r_{16}=$ & 0 & 0 & 1 & 0 & 0 &0 & $\ra 1+0$ \\ \hline
$r_{17}=$ & 0 & 0 & 0 & 0 & 1 & 0 & $\ra 1+0$ \\ \hline
$r_{18}=$ & 0 & 0 & 0 & 1 & 0 & 0 & $\ra 1+0$ \\ \hline
&  &  &  &  &  &  & \\ \hline
$r_{19}=$ & 0 &0 & 0 & 0 & 0 & 2 & $\ra 1+1$ \\ \hline
&  &  &  &  &  &  & \\ \hline
$r_{20}=$ & 0 & 0 & 0 & 0 & 0 & 0 & $\ra 0+1$ \\ \hline
\end{tabular}

{\sl Table 5: The workings of Vertical Upgrade A}

In our example  the spectrum $v_1<v_2<\cdots<v_t$ from 5.1 becomes $2<4<8<16<32<64$. It is easy, yet tedious by hand, to verify

$\Nv=\Nv[\ge 2]=r_1\uplus\cdots\uplus r_4,\ 
\Nv[\ge 4]=r_5\uplus\cdots \uplus r_9,\ \Nv[\ge 8]=r_{10}\uplus\cdots \uplus r_{14},$
$\Nv[\ge 16]=r_{15}\uplus\cdots \uplus r_{18},\ \Nv[\ge 32]=r_{19},\ \Nv[\ge 64]=r_{20}.$

For instance $\{1,2,6\}\in r_3$ and  $|T_1\cap T_2\cap T_6|=2$. Hence
$\{1,2,6\}\in \Nv[\ge 2]\setminus\Nv[\ge 4]$.  The expression $a+b$ at the end of each row in Table 5 shows how many sets in that row have odd and even cardinality respectively. For instance $r_5$ contains $a=3$ sets of odd and $b=4$ sets of even cardinality. It e.g. follows that $|\Nv[\ge 4]'|=3+1+1+1+0=6$ and $|\Nv[\ge 4]''|=4+1+1+1+1=8$.
Similarly 
$|\Nv[\ge 8]'|=6,\ |\Nv[\ge 8]''|=3$.
We deduce from (16) and (20) that

\begin{description}
\item{ } $N[4]'=|\Nv[\ge 4]'| -  |\Nv[\ge 8]'| = 6-6={\bf 0}$
\item{}   $N[4]''=|\Nv[\ge 4]''| -  |\Nv[\ge 8]''| = 8-3={\bf 5}$
\end{description}

Likewise one calculates

$N[2]'=N[2]''=1,\ N[8]'=N[8]''=2,\ N[16]'=3,$

$ N[16]''=0,\  N[32]'=1,\ N[32]''=0,\ N[64]'=0,\ N[64]''=1.$

It now follows from (17) that

$N=\Big(2\cdot 1+4\cdot{\bf 5}+8\cdot 2+16\cdot 0+32\cdot 0+64\cdot 1\Big)
-\Big(2\cdot 1+4\cdot{\bf 0}+8\cdot 2+16\cdot 3+32\cdot 1+64\cdot 0\Big)=4.$

For this small example one verifies ad hoc that indeed $|Mod(\varphi)|=|(2,0,1,1,2,1)|=4$.

{\bf 5.3} Returning to (18), how are we to sieve $\Nv[\ge v_{k+1}]$ from $\Nv[\ge v_k]$ in general?
In a nutshell, starting with ${\cal S}:=\Nv[\ge v_k]$ we keep on removing {\it bad} faces $Y$ (i.e. $N(\ul{Y})=v_k$) from the shrinking set ${\cal S}$ until ${\cal S}=\Nv[\ge v_{k+1}]$. Our particular way of removing $Y$ from ${\cal S}$ is such that along with $Y$ all supersets $Z$ in the same 012n-row as $Y$ get also removed. This is just as well since all these $Z$ are necessarily bad themselves: $Y\subseteq Z$ implies $N(\ul{Z})\le v_k$, yet $N(\ul{Z})<v_k$ is impossible in view of $Z\in\Nv[\ge v_k]$. In fact in order to remove a lot of sets $Z$ it pays to find {\it minimal} bad faces $Y$; how to find them is explained in 5.3.2. While ${\cal S}$ shrinks, old 012n-rows get replaced by new ones. As soon as no 012n-row contains any minimal bad face $Y$, the (disjoint) union of all current 012n-rows is exactly $\Nv[\ge v_{k+1}]$.

{\bf 5.3.1} To fix ideas, suppose $Y_1=\{4,5,6\}\in r_1$ is a minimal bad face of $\Nv[\ge v_k]=r_1\uplus r_2$, where $r_1,\ r_2$ are as in Table 6. Removing $Y$ and all its supersets from $r_1$ results in a set system that can  be represented as $r_3\uplus r_4\uplus r_5$. (See [W2] for details on the $n$-algorithm.) Suppose $r_3$ and $r_4$ do not contain any minimal bad faces (and thus no bad faces at all) but $r_5$ contains the minimal bad face $Y_2=\{4,5,8,9\}$. Suppose after its removal (replace $r_5$ by $r_6$) there are no minimal bad faces left in $r_6$. Therefore we turn to $r_2$. Suppose $Y_3=\{2,3,4,5\}$ is the only minimal bad face contained in $r_2$. Upon removing it we have achieved the representation $\Nv[\ge v_{k+1}]= r_3\uplus r_4\uplus r_6\uplus r_7\uplus r_8\uplus r_9.$

\begin{tabular}{l|c|c|c|c|c|c|c|c|c|c}
&  1 & 2& 3 & 4 & 5 & 6 & 7 & 8 & 9\\ \hline
$r_1 =$ & 0 & $n_1$ & $n_1$ & $n_1$ & 2 & $n_2$ & $n_2$ & $n_2$ & $n_2$ \\ \hline
$r_2 =$ & 1 & $n_1$ & $n_2$ & $n_2$ &  $n_3$ & $n_3$ & $n_3$ & $n_1$ & $n_2$ \\ \hline
 &  &  &  &  &  &  & & & \\ \hline
$r_3 =$ & 0 & 2 & 2 & {\bf 0} & 2 & $n_2$ & $n_2$ & $n_2$ & $n_2$ \\ \hline
$r_4 =$ & 0 & $n_1$ & $n_1$ & {\bf 1} & {\bf 0} & $n_2$ & $n_2$ & $n_2$ & $n_2$ \\ \hline
$r_5 =$ & 0 & $n_1$ & $n_1$ & {\bf 1} & {\bf 1} & {\bf 0} & 2 & 2 & 2 \\ \hline
$r_2 =$ & 1 & $n_1$ & $n_2$ & $n_2$ &  $n_3$ & $n_3$ & $n_3$ & $n_1$ & $n_2$ \\ \hline
 &  &  &  &  &  &  & & & \\ \hline
$r_3 =$ & 0 & 2 & 2 & 0 & 2 & $n_2$ & $n_2$ & $n_2$ & $n_2$ \\ \hline
$r_4 =$ & 0 & $n_1$ & $n_1$ & 1 & 0 & $n_2$ & $n_2$ & $n_2$ & $n_2$ \\ \hline
$r_6 =$ & 0 & $n_1$ & $n_1$ & 1 & 1 & 0 & 2 & ${\bf n_2}$ & ${\bf n_2}$ \\ \hline
$r_2 =$ & 1 & $n_1$ & $n_2$ & $n_2$ &  $n_3$ & $n_3$ & $n_3$ & $n_1$ & $n_2$ \\ \hline
 &  &  &  &  &  &  & & & \\ \hline
$r_3 =$ & 0 & 2 & 2 & 0 & 2 & $n_2$ & $n_2$ & $n_2$ & $n_2$ \\ \hline
$r_4 =$ & 0 & $n_1$ & $n_1$ & 1 & 0 & $n_2$ & $n_2$ & $n_2$ & $n_2$ \\ \hline
$r_6 =$ & 0 & $n_1$ & $n_1$ & 1 & 1 & 0 & 2 & $n_2$ & $n_2$ \\ \hline
$r_7 =$ & 1 & {\bf 0} & $n_2$ & $n_2$ & $n_3$ & $n_3$ & $n_3$ & 2 & $n_2$ \\ \hline
$r_8 =$ & 1 & {\bf 1} & ${\bf n_2}$ & ${\bf n_2}$ & $n_3$ & $n_3$ & $n_3$ & 0 & 2 \\ \hline
$r_9 =$ & 1 & {\bf 1} & {\bf 1} & {\bf 1} & {\bf 0} & 2 & 2 & 0 & 0 \\ \hline
\end{tabular}

{\sl Table 6: Some technicalities of Vertical Upgrade A}

{\bf 5.3.2} It remains to see how the minimal bad faces $Z\in\Nv[\ge v_k]$ can be found, i.e. $Z$ has $N(\ul{Z})=v_k$ but is such that $Z_0=Z\setminus\{z\}$ has $N(\ul{Z_0})>v_k$ for all $z\in Z$. For starters, once we get a hold of any bad face $Y\in r\subseteq \Nv[\ge v_k]$ we simply keep on removing random elements of $Y$ until we arrive at a {\it minimal} bad face $Z\subseteq Y$. (Notice that $Z$ may be located in {\it another} 012n-row $r'\subseteq
\Nv[\ge v_k]$.) 

 But how to find {\it any} bad face in one of the 012n-rows $r$ constituting $\Nv[\ge v_k]$? Since each bad face is contained in a $r$-maximal face, which itself is necessarily bad, it suffices to scan the $r$-maximal faces. If $r$ has $t$ many $n$-wildcards of lengths $\nu_1,\ldots,\nu_t$ respectively, then  $r$ has $\nu_1\nu_2\cdots\nu_t$ many $r$-maximal members. For instance, the $r$-maximal members of $r$ in Table 7 are the $\nu_1\nu_2={3\choose 2}{4\choose 3}=12$ sets $\{4,5,7\}\uplus A\uplus B$ where $A$ ranges over the 2-element subsets of $\{1,2,3\}$ and $B$ ranges over the 3-element subsets of $\{8,9,10,11\}$.

\begin{tabular}{l|c|c|c|c|c|c|c|c|c|c|c|c}
&  1 & 2& 3 & 4 & 5 & 6 & 7 & 8 & 9 & 10 &11\\ \hline
$r =$ & $n_1$ & $n_1$ & $n_1$ & 2 & 1 & 0 & 1 & $n_2$ & $n_2$ & $n_2$ & $n_2$\\ \hline
\end{tabular}

{\sl Table 7: The $r$-maximal members are easily found.}

 Numerical experiments will  be included in a forthcoming version of this article. They will further improve the 
Upgrade B in [W3] that was applied to the same kind of problem.

\section*{References}
\begin{enumerate}
\item[{[NW]}] D.Q. Naiman, H.P. Wynn, Inclusion-exclusion-Bonferroni identities and inequalities for discrete tube-like problems via Euler characteristics, The Annals of Statistics 20 (1992) 43-76.
\item[{[W1]}] M. Wild, A novel type of branch and bound for maximum independent set, arXiv Feb 2010. (An improved version of this draft is in preparation.)
	\item[{[W2]}] M. Wild, Compactly generating all satisfying truth assignments of a Horn formula, Journal on Satisfiability, Boolean Modeling and Computation 8 (2012) 63-82.
	\item[{[W3]}] M. Wild, Inclusion-exclusion meets exclusion, arXiv, Dec 2013. (Despite the different title this is an older version of the present article.)
	\item[{[W4]}] M. Wild, Counting or producing all fixed cardinality transversals, Algorithmica 69 (2014) 117-129.

\end{enumerate}

\end{document}